\documentclass[12pt]{article}
 \usepackage{epsfig}
 
\begin{document}

\title{WHY ISN'T THE SOLAR CONSTANT A CONSTANT?}

\author{K. J. $LI$\altaffilmark{1,2}, W.  $FENG$\altaffilmark{3}, J. C. $XU$\altaffilmark{1,4},
P. X. Gao\altaffilmark{1},
L. H. $YANG$\altaffilmark{1,4}, H. F. $LIANG$\altaffilmark{5},
L. S. $ZHAN$\altaffilmark{6}}
$^{1}$National
Astronomical Observatories/Yunnan Observatory,
      CAS, Kunming 650011,   China \\
$^{2}$Key Laboratory of Solar Activity, National Astronomical
Observatories, CAS, Beijing 100012, China\\
$^{3}$Research Center of Analysis and Measurement,
Kunming University of Science and Technology, Kunming 650093, China \\
$^{4}$Graduate School of CAS, Beijing 100863, China \\
 $^{5}$Department of Physics, Yunnan Normal University, Kunming 650093,
 China\\
 $^{6}$Jingdezhen Ceramic Institute, Jingdezhen 333001, Jiangxi,
China\\


\begin{abstract}
In order to probe the mechanism of variations of the Solar Constant on the inter-solar-cycle scale,
 total solar irradiance (TSI, the so-called Solar Constant) in the time interval of 7 November 1978 to 20 September 2010 is decomposed into three components through  the empirical mode decomposition and time-frequency analyses.
The first component is the rotation signal, counting up to $42.31\%$ of the total variation of TSI, which is understood to be mainly caused by large magnetic structures, including sunspot groups. The second is an annual-variation signal, counting up to
$15.17\%$ of the total variation, the origin of which is not known at this point in time. Finally, the third is the inter-solar-cycle signal,
counting up to
$42.52\%$, which are inferred to be caused
 by the network magnetic elements in quiet regions, whose magnetic flux ranges from $(4.27-38.01)\times10^{19}$ Mx.
\end{abstract}


\section{INTRODUCTION}
The total solar irradiance (TSI) is the total amount of solar electromagnetic energy over the entire spectrum observed at the top of the Earth's atmosphere per
unit area and per unit time. Before TSI was
measured in space, it was thought to be a constant, due to the low precision of the
ground-based instruments at that time, and it was consequently known as the ¡°Solar Constant¡±,
having a value of about 1366 W m$^{-2}$ (Passos et al. 2007; Fr\"ohlich 2009). At present, the Solar Constant is known to vary on all time scales
at which it has been measured, i.e.
minutes to decades (Willson $\&$ Hudson 1991; Fr\"ohlich 2009). For example, latest results indicate
a lower value of TSI of 1361 W m$^{-2}$ based on observations during the minimum time of cycle 23 to 24 (Kopp $\&$ Lean
2011).  Irradiance variability on
time scales shorter than one day (minutes to hours) is mainly caused by convection, related
to granulation, mesogranulation, and supergranulation (Wolff $\&$ Hickey 1987; Solanki et al. 2003). Short-term changes of TSI on time scales of a
few days to weeks are dominated by magnetic structures (Chapman 1987; Solanki et al. 2003). Over the solar cycle, TSI variations of about
$0.1\%$ are thought to come mainly from the combination of the sunspots¡¯ blocking and the intensification due to
bright faculae, plages, and network elements, with a slight dominance of the bright-feature
effect during the time of the maximum of a Schwabe solar cycle (Hudson et al. 1982; Pap et al. 1990). Space-based observations have existed for only about 30 years; therefore, variations
on time scales longer than the Schwable cycle are could not yet be measured directly (Mekaoui $\&$ Dewitte 2008; Li et al. 2010).

The variation of TSI has important implications for our understanding of solar internal
structure, global changes in the Earth's climate system, and solar-terrestrial relationships
(Egorova et al. 2005; Dameris et al. 2006; Krivova  et al. 2007; Krivova $\&$ Solanki 2008; Gan $\&$ Li 2010; Li et al. 2010). Since the coupled system of the Earth's atmosphere and oceans reacts rather
slowly to the varying solar signal, variations of solar irradiance on long time scales are
possibly of even greater importance for the global
climate change (Solanki $\&$ Schussler 2000; Domingo et al. 2009). In this paper, we investigate the reason to what extent various components add to TSI variations on the scale of longer than 1 day, namely why the so-called Solar Constant isn't a constant, through analyzing  a directly measured time series of TSI, the PMOD composite of TSI from 7 November 1978 to 20 September 2010 by the empirical mode decomposition (EMD) method.

\section{INVESTIGATION OF VARIATIONS OF DAILY TSI}
\subsection{DATA}
The PMOD composite is an accurate measurement record of daily TSI during the last three solar cycles (Fr\"ohlich 2006, 2009). Daily TSI from 7 November 1978 to 20 September 2010 in the PMOD composite is used here to
investigate to what extent TSI varies on various time scales.
The time series can be downloaded from the web site\footnote{ftp://ftp.pmodwrc.ch/pub/data/irradiance/composite/}.
Figure 1  shows the PMOD composite of daily TSI.
The two most striking features of the observed record of the daily TSI (for details, see Fr\"ohlich 2009) are the inter-solar-cycle variations by about  $0.1\%$ in phase with the solar activity cycle and sharp dips with a comparable or even greater amplitude typically lasting 7 to 10 days (Krivova $\&$ Solanki 2008). The mean value of the composite is 1365.91 W m$^{-2}$ during the time interval considered.

\subsection{EMD ANALYSIS OF DAILY TSI}
The empirical mode decomposition (EMD) is a nonlinear time-frequency
analysis method (Huang et al. 1998; Gao et al. 2011). It is an algorithm which decomposes an input signal
into a finite set of oscillating functions, namely
the so-called intrinsic mode functions (IMFs),
which are the intrinsic periodicities of the original signal.
These IMFs are extracted from the data themselves, and they are not restricted to have
constant phases or amplitudes. Essentially EMD is an empirical algorithm which decomposes a signal, which
can be non-stationary and nonlinear, into a finite set of IMFs  (Barnhart $\&$ Eichinger 2011). These IMFs are defined to be functions which are symmetric about their local mean, and whose number of extrema and zero-crossings are equal or differ at most by one (Huang et al. 1998). These IMFs are extracted from a signal using a process called sifting.
 The sifting process essentially iteratively removes the local mean from a signal to extract
the various cycles present. The sifting process is performed until the signal meets the definition
of an IMF (for details, please see Barnhart $\&$ Eichinger (2011)). Here,
the PMOD composite  is decomposed into 10 IMFs through the EMD analysis, which are shown
in Figure  2.

The code of the wavelet transform analysis, which is provided by Torrence $\&$ Compo (1998), is utilized to study  periodicity in the first 9 IMFs of the PMOD composite. IMF 10 is excluded, because it is the secular trend of the composite, and limited length of the data used gives no period to the trend at present.
Figure 3 shows their global wavelet power spectra and the corresponding $95\%$ confidence level. Table 1 gives the periods in the first 9 IMFs, which are significant at
the $95\%$ confidence level. Also given in the table are the period intervals of these period values.

\begin{center}
\begin{table}
 \caption{The periods (in days) in the first 9 IMFs of daily TSI}
 \scriptsize{
 \begin{tabular}{lllllllll}
 \hline
      IMF 1 & IMF 2 & IMF 3 & IMF 4 & IMF 5 & IMF 6 & IMF 7 & IMF 8 & IMF 9 \\
 \hline
 $9.8\pm0.7$  & $14.5\pm1.1$ & $29.0\pm2.3$  & $58.1\pm4.3$ &  $86.7\pm6.1$ & $376.0\pm16.3$  & $390.6\pm22.3$  & $781.1\pm43.2$  & $1104.7\pm52.1$ \\
           &      &      &      &       &       & $781.1\pm43.2$ & $1570.0\pm67.5$ & $3880.2\pm127.5$ \\
  \hline
 \end{tabular}}
 \end{table}
\end{center}

 The periods,  9.8,  14.5, 58.1, and 86.7 days are inferred to be the 1/3-,  1/2-, 2-, and 3-multiple harmonics  of the period of about 29 days,  which is approximately the solar rotation period. There are only those periods in IMFs 1 to 5, which are related to the rotation cycle,  thus IMFs 1 to 5 are called the rotation-variation signal of TSI.
  The sum of IMFs 1 to 5, which is called here Component I of daily TSI, is shown in Figure 4.
 Component I is inferred to be mainly caused by magnetic structures, including sunspot groups, due to the following
 aspects. (1) Short-term changes of TSI on time scales of a
 few days to weeks are known to be dominated by magnetic structures (Chapman 1987; Solanki et al. 2003).
   (2) The figure shows that Component I fluctuates with much higher amplitude around the maximum times of the Schwable cycles than around the minimum times, and long-lived solar magnetic structures usually appear around the maximum times of solar cycles. And (3) as the figure displays,
    sharp dips  appear only in this component and around the maximum times of solar cycles, and maximum variation amplitude can even exceed  3 W m$^{-2}$. Here, the variation amplitude is given relative to the mean value of the composite. These dips, lasting 7 to 10 days,  are caused by the passage of sunspot groups across the visible disk as the Sun rotates. Towards activity maxima, when the number of sunspots grows considerably, the frequency and depth of the dips increase (Krivova $\&$ Solanki 2008). Figure 5 shows daily sunspot area  and daily TSI  from 2003 September 10 to 2003 November 17. The figure displays the very well-known fact that when large sunspot groups pass across the solar visible disk, a sharp dip appears in the daily TSI with its amplitude decreasing from about {\it 1366 W m$^{-2}$} to about {\it 1362 W m$^{-2}$}.  These sharp dips of TSI are caused by the passage of sunspot groups across the visible disk as the Sun rotates.

 The periods, 781.1 and 1570 days are
considered to be the 2- and 4-multiple harmonics  of the period of 390.6 days, respectively, and these three periods show a broad peak in their power spectra.
 Thus, IMFs 6 to 8 show periodical annual variations, and they are called the annual-variation signal of TSI.

 Periodical annual variation  signal  had not been
 determined  by the helioseismic probing of the  solar interior
  (Howe et al. 2000). {\it The one-year periodicity
is found in several solar-activity indices, but its origin is doubtful. That is, it is difficult to
rule out the possibility that this periodicity is not due to the influence of seasonal effects}
  (Javaraiah et al. 2009). Of course, it must be pointed out that, so far there has been no quantitative analysis about effect of the Earth's helio-latitude on the measurement of the Sun, and the origin of the annual periodical signal of the Sun is an open issue.
 Here, we speculate that IMFs 6 to 8 are possibly caused by the
Earth's orbital revolution. However this needs to be independently
confirmed through the analysis of the Earth's/spacecraft orbital data
along with the TSI time series.
 The sum of IMFs 6 to 8, which is called here Component II of daily TSI, is shown in Figure 4, and  almost all variation amplitudes are found less than  0.5 W m$^{-2}$.

 We also calculate the correlation coefficient ($cc$) between daily TSI of the PMOD composite and daily sunspot number, which is available from Solar Influences Data Analysis Center's (SIDC) web site, and $cc=0.4456$, which is statistically significant at the $99.9\%$ confidential level. When
the annual-variation signal of daily TSI, namely Component II is deduced from the original daily TSI, $cc$ obviously increases to be 0.4659. Based on the method used to test the statistical difference of two correlation coefficients by Li et al. (2002), the difference between these two $cc$ values is found significant with a probability of about $91\%$, that is to say, the difference is not caused by randomness, and the two values are statistically different from each other.

\subsection{RELATION BETWEEN THE SCHWABE-CYCLE-RELATED COMPONENT OF DAILY TSI AND MAGNETIC ACTIVITY}
 The period of 3880.2 days ($\approx 10.63$ years) corresponds to the so-called Schwabe cycle, and the period, 1104.7 days is inferred to be the 3-multiple harmonic of the approximately annual period of about 376.0 days. IMF 9, probably plus IMF 10 is therefore related with magnetic activity of the Schwabe cycle.

 Jin et al (2011) have divided full-Sun's magnetograms into active regions (AR) and quiet regions (QR) and calculated the monthly average magnetic flux $F_{{\rm AR}}$ and $F_{{\rm QR}}$ respectively in the time interval of September 1996 to February 2010 with the MDI/SOHO data used.
 They found that the flux of network magnetic elements in QR
 could be further divided into four components: (1) those elements, whose fluxes are
in the  range of $(1.5-2.9)\times10^{18}$ Mx,  are basically independent
of the sunspot cycle, and thus called by them {\it no-correlation} elements ($F_{{\rm no}}$);
 (2) the elements in the flux range of $(2.9-35.9)\times10^{18}$ Mx show an in-phase correlation with
  the sunspot cycle, and thus they are
  {\it anti-phase} elements ($F_{{\rm anti}}$);
(3) those in the flux range of $(35.9-42.7)\times10^{18}$ Mx are called {\it transition} elements ($F_{{\rm tran}}$),  which represents a transition from
anti-phase to in-phase with the sunspot cycle; and
(4) the so-called {\it in-phase}  elements ($F_{{\rm in}}$), in the range of $(4.27-38.01)\times10^{19}$ Mx, which is in-phase with the sunspot cycle.  Based on IMFs 9 and 10, we calculate the monthly average value ($\overline{IMF_{9}}$) of IMF 9 and that
($\overline{IMF_{9+10}}$) of IMF 9 plus IMF 10 in the time interval of September 1996 to February 2010. Then
we calculate the correlation coefficient of $\overline{IMF_{9}}$ and $\overline{IMF_{9+10}}$ respectively
 with $F_{{\rm AR}}$, $F_{{\rm QR}}$, $F_{{\rm no}}$, $F_{{\rm anti}}$, $F_{{\rm tran}}$,  and $F_{{\rm in}}$,  and the results obtained are
 given in Table 2.

\begin{center}
\begin{table}
\begin{center}
 \caption{Correlation coefficients of TSI components  with magnetic activity}
 \begin{tabular}{cccccccccc}
 \hline
            & $F_{{\rm AR}}$& $F_{{\rm QR}}$& $F_{{\rm no}}$& $F_{{\rm anti}}$& $F_{{\rm tran}}$& $F_{{\rm in}}$ \\
$\overline{IMF_{9}}$   & 0.8319 &  0.6737  & -0.4891 & -0.6991   & -0.3991   & 0.7198\\
$\overline{IMF_{9+10}}$& 0.9671 &  0.9518  & -0.0046 & -0.5804   &  0.0824   & 0.9818\\
  \hline
 \end{tabular}
\end{center}
 \end{table}
\end{center}

The relation of $\overline{IMF_{9+10}}$ respectively with $F_{{\rm in}}$ and $F_{{\rm AR}}$ gives
the maximum two correlation coefficients among these coefficients in the table, which are
correspondingly much larger than those given by the relations of $\overline{IMF_{9}}$ respectively
with $F_{{\rm in}}$ and $F_{{\rm AR}}$, and the maximum correlation coefficient is given for the relation of
$\overline{IMF_{9+10}}$ with $F_{{\rm in}}$. Thus, it is seemingly $\overline{IMF_{9+10}}$, not $\overline{IMF_{9}}$ that is most probably related
with magnetic activity, and the magnetic activity is referred to the magnetic flux of  $F_{{\rm in}}$.
That is to say, IMF 9 plus IMF 10 should be related with magnetic activity
of the Schwabe cycle.
Figure 6 plots  $\overline{IMF_{9}}$ and $\overline{IMF_{9+10}}$ together with the
monthly average magnetic flux values of $F_{{\rm in}}$ and $F_{{\rm AR}}$, in order
to illustrate the relations of $\overline{IMF_{9}}$ and $\overline{IMF_{9+10}}$ respectively with $F_{{\rm in}}$ and $F_{{\rm AR}}$.
 The sum of IMFs 9 and 10, which is called here Component III of daily TSI, is shown in Figure 4. Variation amplitude of Component III  can reach  0.6 W m$^{-2}$, and the difference of its maximum and minimum values
can still match up to TSI variations of about $0.1\%$.

In order to examine the significance in the difference of the maximum two
coefficients in the above table, a statistical test is carried out following Li et al (2002),
and then the difference  is found significant with a probability of about
$97\%$. Thus, Component III  is inferred to be caused
 by the network magnetic elements, whose magnetic flux are of $(4.27-38.01)\times10^{19}$ Mx.
 The above significant difference somewhat confirms that magnetic fields of different strengths could even act on TSI in reverse ways:
 intense magnetic fields, as thermal 'plugs' to divert heat flow from solar deep layers, decrease TSI, but small-scale
 magnetic fields, as local thermal 'leaks', increase TSI (Domingo et al. 2009).

The complex Morlet  wavelet transform  is utilized to study the periodicity respectively in Components I and II.
Figure 7 shows their global wavelet power spectra and corresponding $95\%$ confidence levels.
For Component I the periods of significance are $14.2\pm1.1$  and $31.7\pm2.8$ days, and for Component II, $366\pm15.0$ and $726.3\pm38.4$ days, which are all significant at the $95\%$ confidence level. Thus, Component I is indeed the rotation signal of TSI, and Component II, the annual-variation signal.

Finally, we determine the contribution of each of the three components to the daily TSI. We calculate the sum of Components I, II, and III  over the whole time interval, respectively. The sum of Component I counts up to $42.31\%$ of the
total sum of Components I to III, namely daily TSI related to its mean value (TSI minus its mean value). The sum of Component II counts up to  $15.17\%$, and the sum of Component III, $42.52\%$.

\section{CONCLUSIONS}
 Firstly, the PMOD composite of daily TSI in the time interval of 7 November 1978 to 20 September 2010
 is decomposed into 10 intrinsic mode functions (IMFs) through the empirical mode decomposition analysis.
 Secondly, the  Morlet  wavelet transform  is utilized to study  periodicity in the first 9 IMFs (the 10th IMF shows the secular trend of TSI). And lastly, correlation analyses of IMF 9 and IMF 9 plus IMF 10 are made respectively with the magnetic flux of active regions, that of quiet regions, and  that of network elements in quiet regions with different magnetic fluxes.  Resultantly, a new mechanism is proposed to explain why the Solar Constant isn't a constant. The main conclusions are obtained as follows.

Daily TSI is found to mainly consist of three components. The first one is the rotation signal, counting up to
$42.31\%$ of the total variation of TSI, which is inferred to be mainly caused by large magnetic structures, including sunspot groups. The second is the annual-variation signal, counting up to
$15.17\%$ of the total variation.
We speculate that it is caused by the annual change of the Earth's helio latitude.
It should be pointed out here that we do not give a quantitative analysis about the effect of the Earth's helio-latitude on the measurement of the Sun, for that we do not know the detailed latitudinal distribution of TSI on the solar disk.
The origin of the annual periodical signal of the Sun is an open issue, and further research is needed in future.
And the third is the signal varying at the scale of the Schwabe cycles,
counting up to
$42.52\%$ of the total variation in TSI, which is inferred to be caused
 by the network magnetic elements in quiet regions, whose magnetic flux are of $(4.27-38.01)\times10^{19}$ Mx.

 When Component II is reduced from the original daily TSI, the reduced daily TSI is more intensely related with daily sunspot number  than the original daily TSI itself is. This leads us
to conclude that Component II should be
caused by the annual change of the Earth's helio-latitude.
\\

Acknowledgments:
We thank the anonymous referees for their careful reading of the
manuscript and constructive comments which improved the original
version of the manuscript.
The authors thank PMOD/WRC,
Davos, Switzerland, for making available the data set of total solar irradiance measurements (version d41-62-
1009), comprising composite irradiance values since 1978, which includes unpublished data from the VIRGO
Experiment on the cooperative ESA/NASA SOHO Mission. This work is supported by the Natural Science
Funds of China (10873032, 10921303, 11147125,  and 11073010), the 973 project (2011CB811406), and the
Chinese Academy of Sciences.

\clearpage

\begin{figure}
\hskip 10.mm
\includegraphics[angle=0,scale=.80]{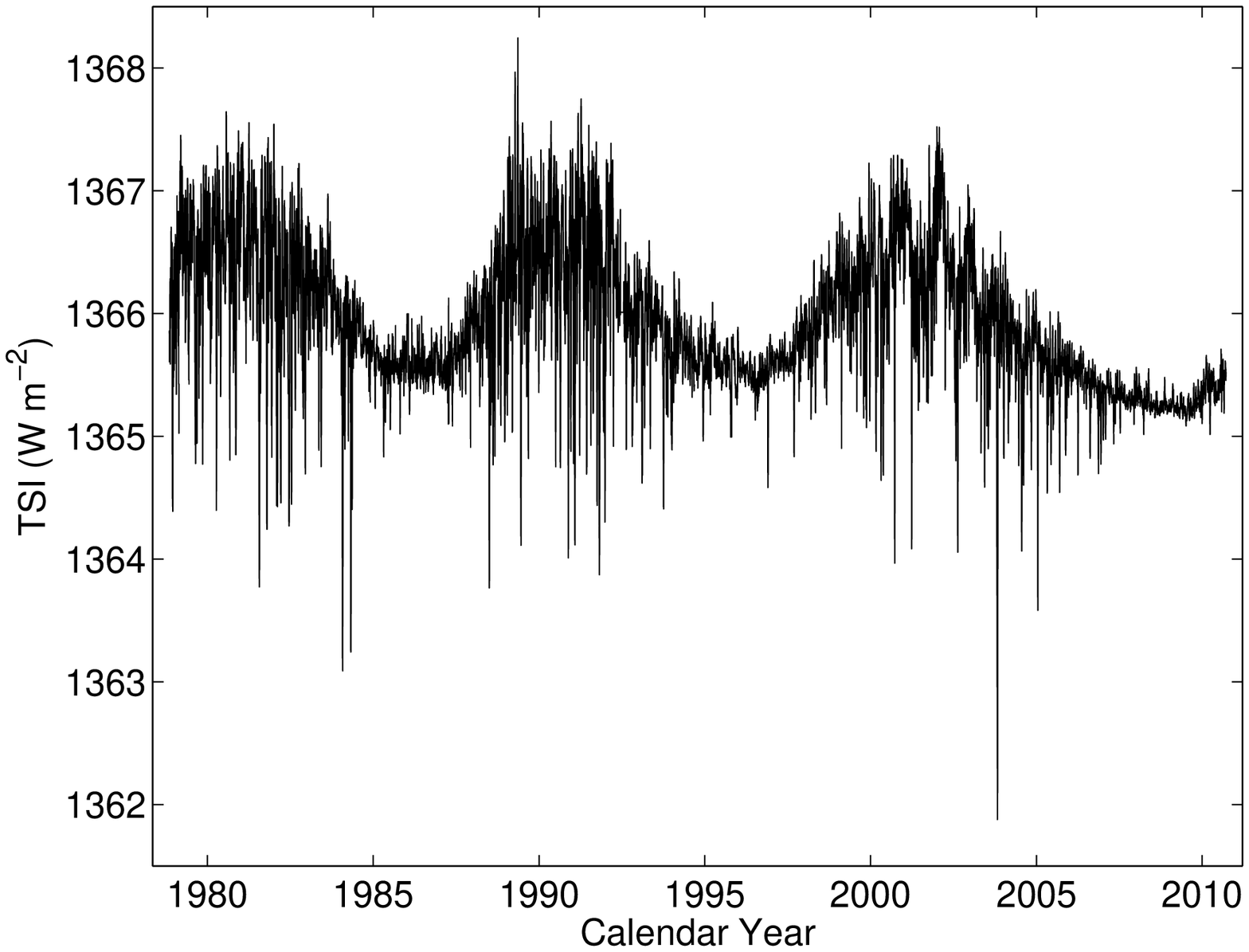}
\caption{The PMOD composite of daily TSI  from 7 November 1978 to 20 September 2010.  }
\end{figure}

\begin{figure}
\hskip 10.mm
\begin{center}
\includegraphics[angle=0,scale=.70]{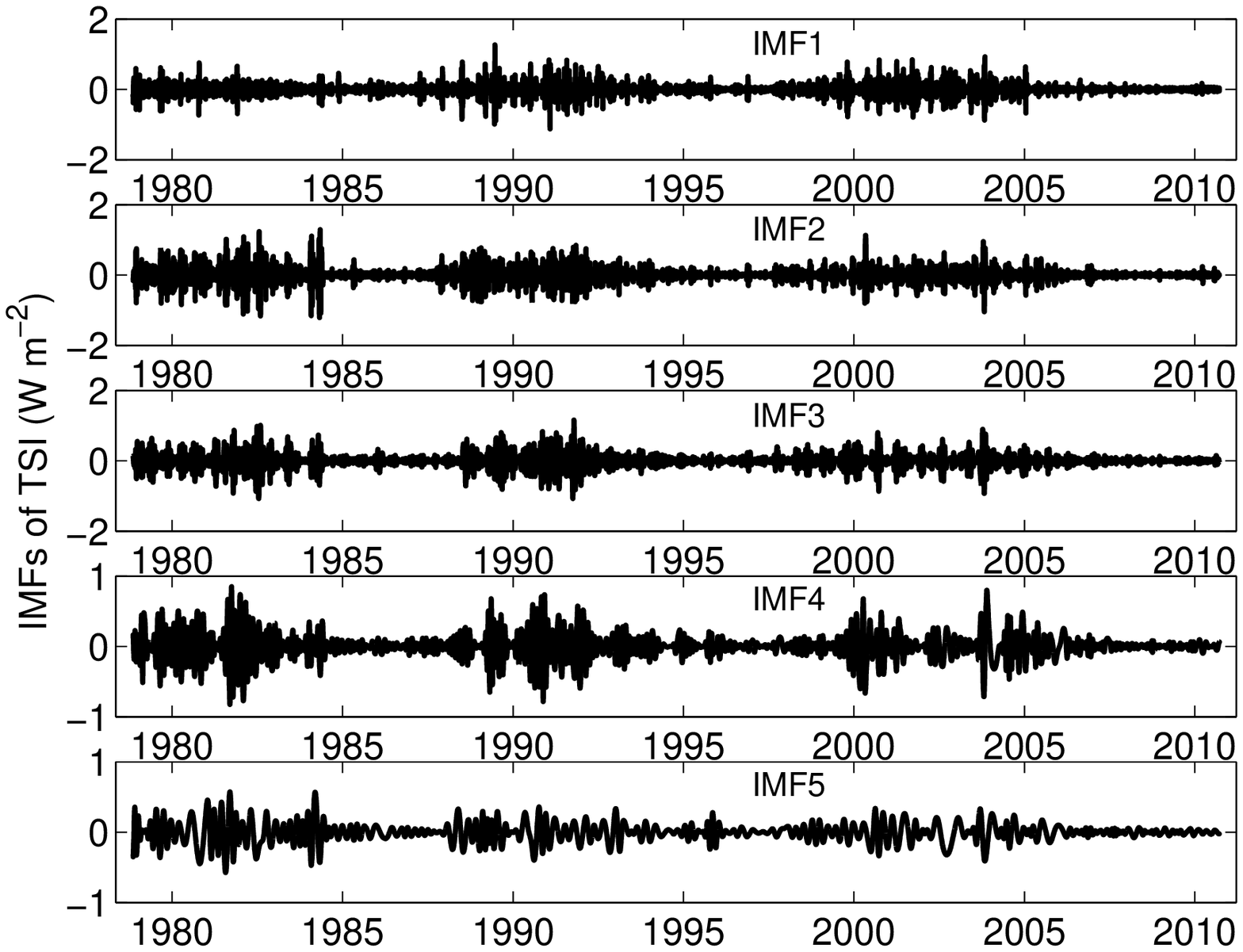}
\includegraphics[angle=0,scale=.70]{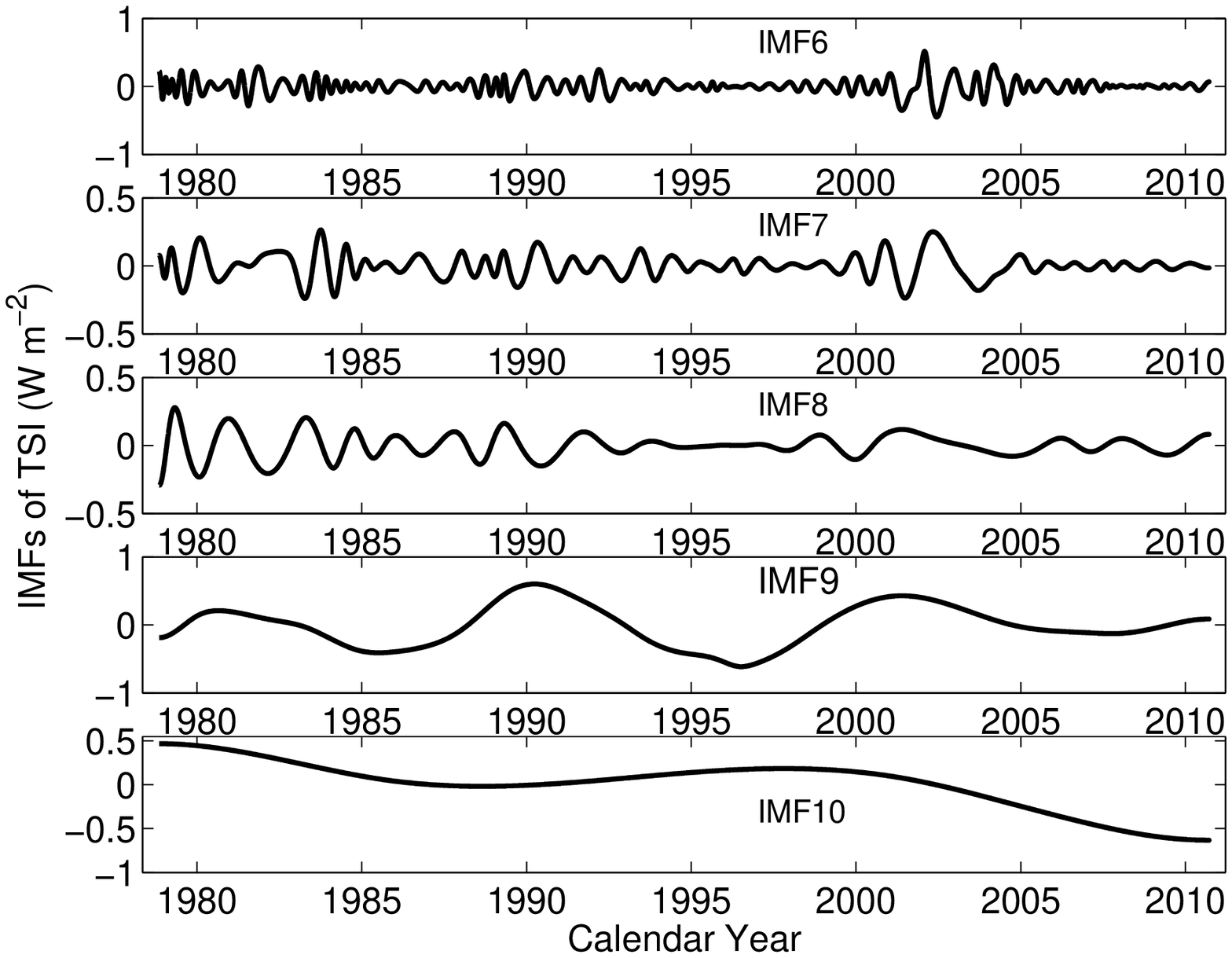}
\caption{The intrinsic mode functions (IMFs) of the PMOD composite.
IMFs 1 to 10 are shown correspondingly in the panels, ranking form the top one to the bottom, respectively.
}
\end{center}
\end{figure}

\begin{figure}
\hskip 10.mm
\begin{center}
\includegraphics[angle=0,scale=.70]{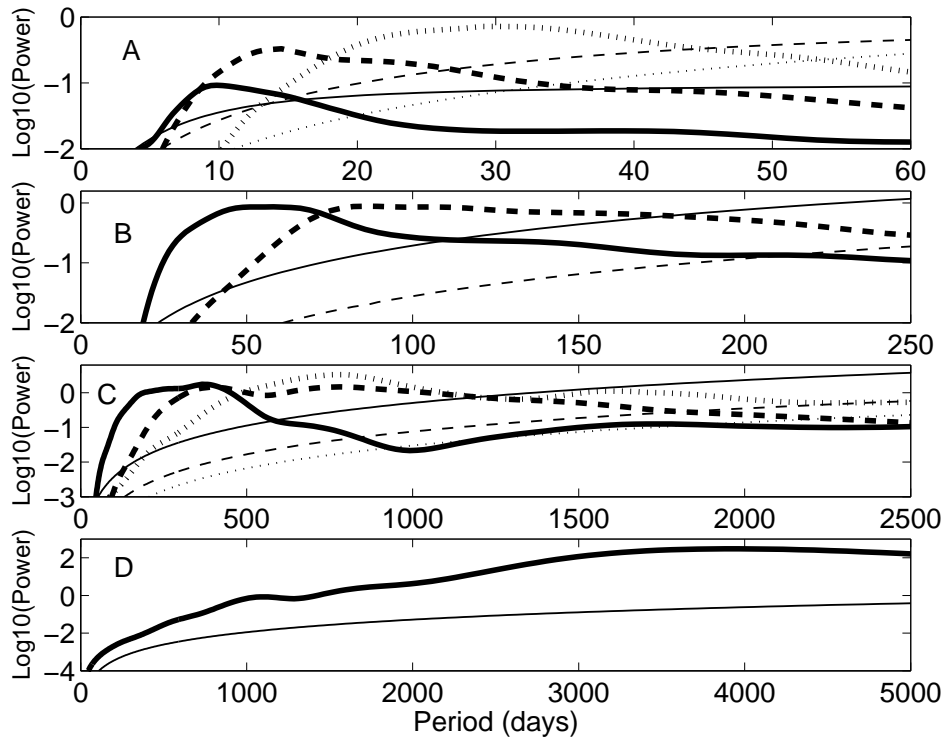}
\caption{The global wavelet power spectra (the thick lines ) of the first 9 IMFs and their corresponding $95\%$ confidence level (the thin lines). Those for IMFs 1, 2, and 3 are shown  in the top panel (Panel A) respectively by
 the solid lines, dashed lines, and dotted lines; for IMFs 4 and 5, in the second panel (Panel B) respectively by
 the solid lines and the dashed lines; for IMFs 6 to 8, in the third panel (Panel C) respectively by
the solid lines , dashed lines, and dotted lines; and for IMF 9, in the bottom panel (Panel D) by the solid lines.
}
\end{center}
\end{figure}

\begin{figure}
\hskip 10.mm
\begin{center}
\includegraphics[angle=0,scale=.70]{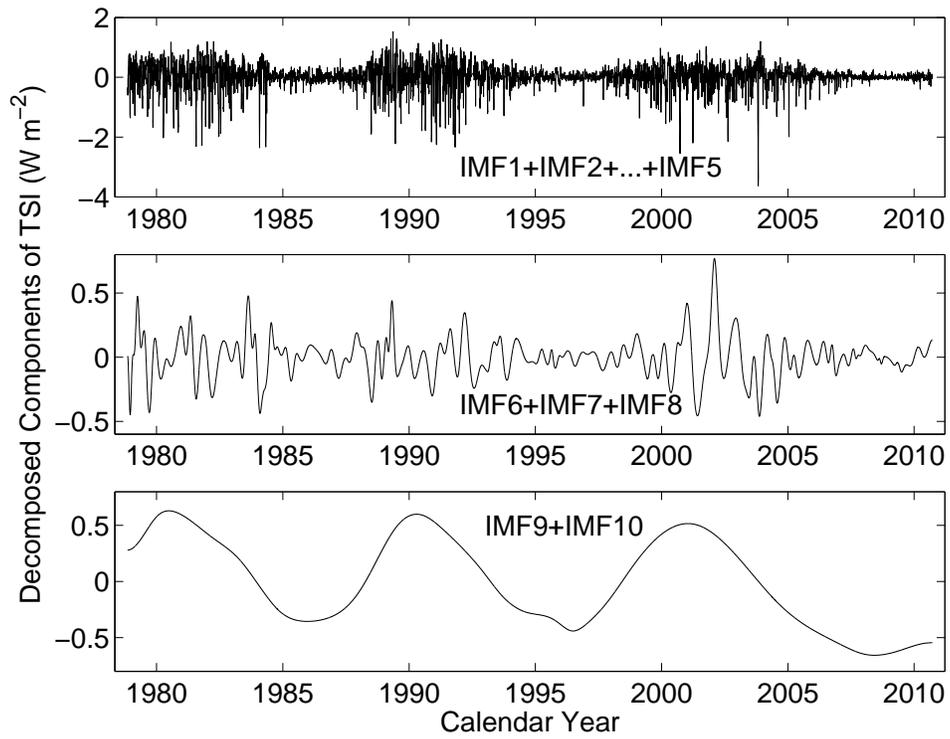}
\caption{Three components of daily TSI. Component I is shown in the top panel, which is the sum of IMFs 1 to 5;
Component II is shown in the middle panel, which is the sum of IMFs 6 to 8; and
Component III is shown in the bottom panel, which is the sum of IMFs 9 to 10.
}
\end{center}
\end{figure}

\begin{figure}
\hskip 10.mm
\begin{center}
\includegraphics[angle=0,scale=.70]{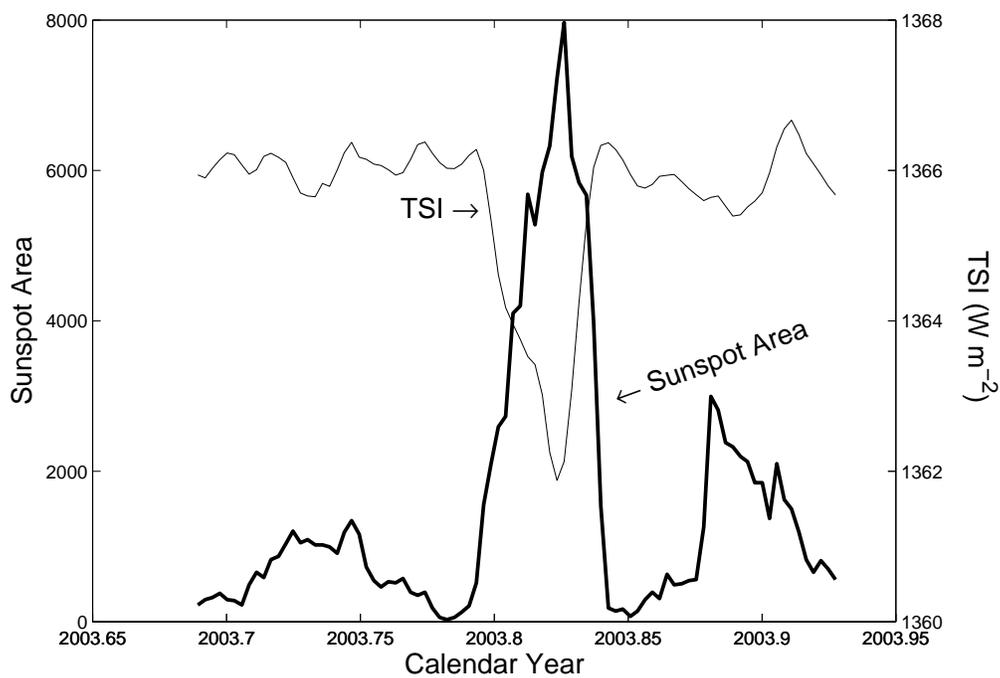}
\caption{Daily sunspot area (the thick line) and daily TSI (the thin line) from 2003 September 10 to 2003 November 17.
}
\end{center}
\end{figure}

\begin{figure}
\hskip 10.mm
\begin{center}
\includegraphics[angle=0,scale=.70]{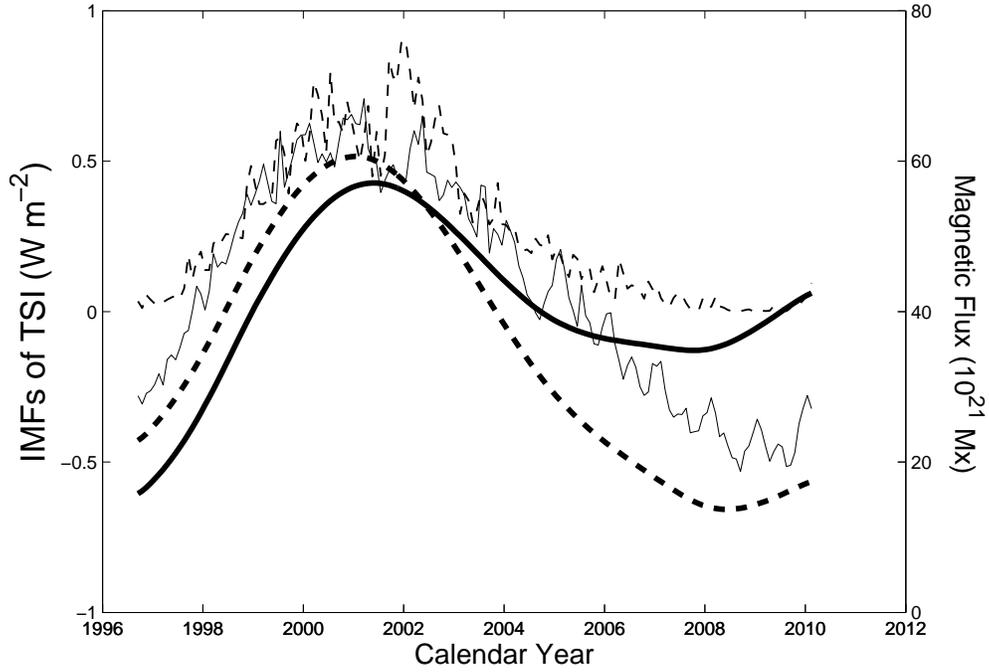}
\caption{Monthly average values  ($\overline{IMF_{9}}$) of IMF 9 (the thick and solid line)
and monthly average values ($\overline{IMF_{9+10}}$) of IMF 9 plus IMF 10 (the thick and dashed line) together with monthly average magnetic flux values of $F_{{\rm in}}$ (the thin and solid line) and $F_{{\rm AR}}$
(the thin and dashed line). $F_{{\rm AR}}$ is divided by a constant, in order to show $F_{{\rm in}}$ together with $F_{{\rm AR}}$ well.
}
\end{center}
\end{figure}

\begin{figure}
\hskip 10.mm
\begin{center}
\includegraphics[angle=0,scale=.70]{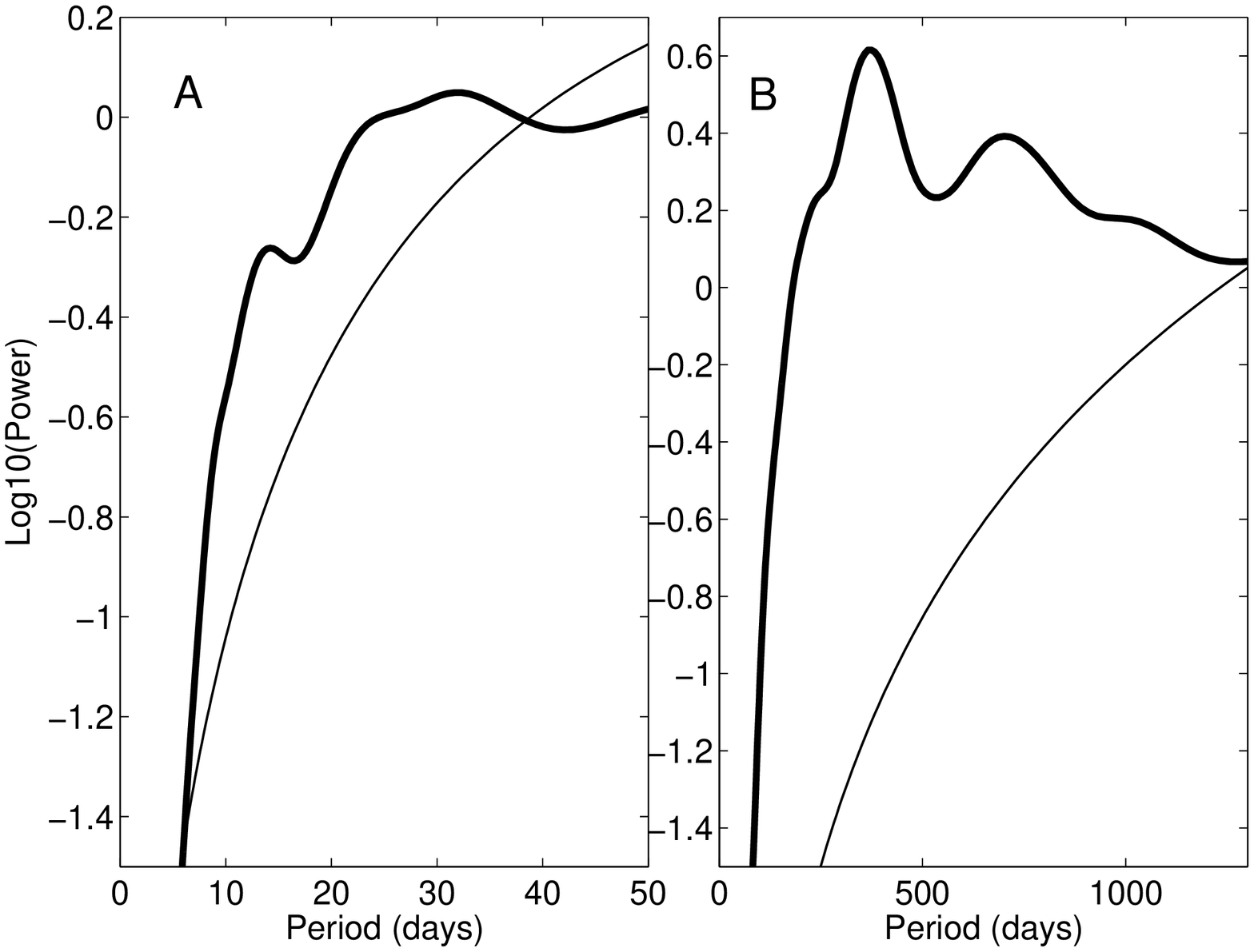}
\caption{The global wavelet power spectra (the thick lines ) of Components I (left panel) and II (right panel).
The thin lines shows their corresponding $95\%$ confidence level.
}
\end{center}
\end{figure}


\begin{thebibliography}{}
\bibitem[Barnhart(2011)]{Bar11}Barnhart, B. L., $\&$ Eichinger, W. E. 2011, Solar Phys., DOI 10.1007/s11207-010-9701-6
\bibitem[Chapman(1987)]{Cha87}Chapman, G. A.  1987,  Annu. Rev. Astron. Astrophys.,   25, 633
\bibitem[Dameris(2006)]{Dam06}Dameris, M., Matthes, S., Deckert, R. Grewe, V., $\&$ Ponater, M. 2006, Geophys. Res. Lett., 33, 3806
\bibitem[Domingo(2009)]{Dom09}Domingo,  V.,  Ermolli, I., Fox,   P., Fr\"ohlich,  C.,  Haberreiter, M., Krivova,  N., Kopp, G., Schmutz, W., Solanki, S. K.,  Spruit, H. C., Unruh,  Y., $\&$ Vogler, A. 2009,  Space Sci. Rev.,  145, 337
\bibitem[Egorova(2005)]{Ego05}Egorova, T., Rozanov,  E., Zubov, V., Schmutz, W., $\&$ Peter, T. 2005,  Adv. Space Res.,   35, 451
\bibitem[Fr\"ohlich(2006)]{Fro06} Fr\"ohlich, C. 2006,  Space Sci. Rev.,   125, 53
\bibitem[Fr\"ohlich(2009)]{Fro09} Fr\"ohlich, C.  2009,  Astron. Astrophys.,   501, L27
\bibitem[Gan(2010)]{Gan10}Gan,  W. Q., $\&$  Li, Y. P. 2010,  Solar and Stellar Variability: Impact on Earth and Planets, Proceedings IAU Symposium, 264, 84
\bibitem[Gao(2011)]{Gao11} Gao, P. X., Liang,  H. F., $\&$ Zhu,  W. W.  2011,    New Astronomy, 16, 147
\bibitem[Howe(2000)]{How00}Howe, R., Christensen-Dalsgaard, J., Hill, F., Komm, R.W., Larsen, R. M., Schou, J., Thompson, M. J., $\&$ Toomre, J. 2000, Science 287, 2456
\bibitem[Huang(1998)]{Hua98} Huang, N. E., Shen,  Z., Long, S. R., Wu,  M. C., Shih,  H. H., Zheng, Q.,  Yen, N. C., Tung, C. C., $\&$
Liu,  H. H. 1998,   Proceedings of Royal Society London, Ser. A,   454, 903
\bibitem[Hudson(1982)]{Hus82}Hudson, H. S.,  Silva, S., Woodard, M., $\&$ Willson, R. C., 1982,   Sol. Phys.,   76, 211
\bibitem[Javaraiah(2009)]{Jav09}Javaraiah, J., Ulrich, R. K.,  bertello, L., $\&$ Boyden, J. e. 2009, Solll. Phys., 257, 61
\bibitem[Jin(2011)]{Jin11}Jin,   C. L.,  Wang, J. X., $\&$  Song, Q. 2011,  ApJ, 731, 37
\bibitem[Kopp(2011)]{Kop11}Kopp, G., $\&$ Lean J. L. 2011,  Geophys. Res. Lett., 38, L01706, doi:10.1029/2010GL045777
\bibitem[Krivova(2007)]{Kri07}Krivova, N. A., Balmaceda,  L., $\&$  Solanki, S. K. 2007, $A\&A$, 467, 335
\bibitem[Krivova(2008)]{Kri08}Krivova,  N. A., $\&$ Solanki,  S. K. 2008, J. Astrophys. Astron., 29, 151
\bibitem[Li(2002)]{Li02}Li, K. J., Irie, M., Wang, J. X., Xiong, S. Y., Yun, H. S., Liang, H. F., Zhan, L. S., $\&$ Zhao, H. J. 2002, Publ. Astron. Soc. Japan, 54, 787
\bibitem[Li(2010)]{Li10}Li, K. J.,  Xu, J. C.,  Liu, X. H.,   Gao, P. X., Zhan, L. S. 2010, Sol. Phys., 267, 295
\bibitem[Mekaoui(2008)]{Mek08}Mekaoui,  S., $\&$ Dewitte, S. 2008,   Sol. Phys.,   247, 203
\bibitem[Pap(1990)]{Pap90}Pap, J., Tobiska, W. K., $\&$ Bouwer,  S. D. 1990, Sol. Phys.,   129, 165
\bibitem[Passos(2007)]{pas07}Passos D., Brandao, S., $\&$ Lopes, I. 2007, Advances in Space Research, 40, 990
\bibitem[Solanki(2000)]{Sol00} Solanki,  S. K., Schussler,  M., $\&$ Fligge,  M. 2000,  Nature,   408, 445
\bibitem[Solanki(2003)]{Sol03}Solanki, S. K.,  Seleznyov,  A. D., $\&$  Krivova, N. A. 2003, In: Wilson, A. (ed.) International Solar Cycle Studies (ISCS) Symposium,   SP-535, ESA, Noordwijk, 285
\bibitem[Torrence(1998)]{tor98}Torrence, C., $\&$ Compo, G. P. 1998,  Bull.  Amer. Meteor. Soc., 79, 61
\bibitem[Willson(1991)]{Wil91}Willson, R. C., $\&$ Hudson, H. S. 1991,  Nature,   351, 42
\bibitem[Wolff(1987)]{Wolf87}Wolff, C. L.,  $\&$ Hickey, J. R. 1987,   Sol. Phys.,   109, 1
\end{thebibliography}
\end{document}